# Size-dependent properties of dithallium selenide

A. M. Panich[1]*, M. Shao[2], C. L. Teske[3], and W. Bensch[3]

[1] Department of Physics, Ben-Gurion University of the Negev, P.O. Box 653, Beer Sheva 84105, Israel

[2] Anhui Key Laboratory of Functional Molecular Solids, and College of Chemistry and Materials Science, Anhui Normal University, Anhui Province, Wuhu 241000, China

[3] Institute of Inorganic Chemistry, Christian-Albrechts University, Kiel, Germany

* Corresponding author. Fax: +972-8-6472-903, tel.: 972-8-6472-458

e-mail: pan@bgumail.bgu.ac.il

**Abstract**

We report on size-dependent properties of dithallium selenide, $Tl_2Se$. We have carried out a comparative nuclear magnetic resonance (NMR) study of $Tl_2Se$ nanorods and bulk samples, measuring NMR spectra and spin-lattice relaxation rate of $^{203}Tl$ and $^{205}Tl$ isotopes. Though bulk $Tl_2Se$ was reported to be a metal, the Korringa-like spin-lattice relaxation behavior is observed only at low temperatures and is transformed to an activation regime above ~200 K. This finding is interpreted assuming a two-band model in the semimetallic compound. Our measurements show significant difference in the Knight shift and indirect nuclear exchange coupling for the bulk and nanorod $Tl_2Se$ samples, reflecting noticeable distinction in their electronic structure. At that, $Tl_2Se$ nanorods are semiconductors and exhibit a characteristic activation behavior in the spin-lattice relaxation rate due to the thermal excitation of carriers to the conduction band. The obtained size dependence of the $Tl_2Se$ properties is interpreted in terms of the semimetal-semiconductor transformation due to the quantum confinement.

Fabricating and studying metal and semiconductor nanostructures is of great interest, being a promise of applications in modern microelectronics. Reduced dimensionality provides physicists with an opportunity to discover quantum properties that are present neither in the bulk nor at the molecular limit. If a metal particle is reduced in size down to a value corresponding to the de Broglie wavelength, the density of states in the valence and conduction bands decreases and the electronic properties change dramatically. Such an effect may be regarded as the onset of the metal-nonmetal transition. Among many nano-systems, a lot of investigations are focused on quasi-one-dimensional solids such as nanorods and nanowires.

In the present paper, we report on size-dependent properties of dithallium selenide, $Tl_2Se$, which we investigated by means of comparative nuclear magnetic resonance (NMR) studies of $Tl_2Se$ nanorods and bulk samples. We measured NMR spectra and spin-lattice relaxation rates of $^{203}Tl$ and $^{205}Tl$ isotopes, that have spin I=1/2 and are excellent probes in studying of the electronic structure of compounds [1-13]. Our measurements show clear distinction in the indirect nuclear exchange coupling and Knight shift in the bulk and nanorod $Tl_2Se$ samples, reflecting noticeable difference in their electronic structure. Though Mooser and Pearson [14] mentioned that bulk $Tl_2Se$ behaves as a metal, we observed that it exhibits Korringa-like spin-lattice relaxation behavior only at low temperatures. At temperatures higher than ~200 K, it is transformed into an activation regime. The $Tl_2Se$ nanorod sample shows typical activation behavior in the whole temperature range under study, which is characteristic of a semiconductor and is realized due to the thermal excitation of carriers to the conduction band. The obtained size dependence of the $Tl_2Se$ properties is interpreted in terms of the semimetal-semiconductor transformation in $Tl_2Se$ due to the quantum confinement.

All experiments were performed on macroscopic, powder $Tl_2Se$ bulk and nanorod samples. Bulk $Tl_2Se$ was prepared from a stoichiometric mixture of the elements in a glass ampoule (Duran) with argon under reduced pressure at about 400°C. This compound exhibits a layered structure built from triple Tl-Se-Tl sheets that are stacked along the *c*-axis [15]. The structure belongs to the tetragonal symmetry. Metallic-like conductivity of the compound results in enormous reduction of the Q-factor of the tank circuit and does not allow one to measure NMR. Therefore the highly conductive bulk sample was finely ground, sieved through the 53 μm sieve and dispersed in talc. The preparation and characterization of the $Tl_2Se$ nanorods is described elsewhere [16]. The XRD pattern of the $Tl_2Se$ nanorod was assigned to the tetragonal phase; the calculated lattice constants are close to those reported for the bulk compound. The average size of $Tl_2Se$ nanorods is about 75 nm in diameter and 900 nm in length. The nanorods were shown to be semiconductors [16].

$^{203}Tl$ and $^{205}Tl$ NMR measurements of powdered $Tl_2Se$ bulk and nanorod samples were carried out in the temperature range 70 - 295 K using a Tecmag pulse NMR spectrometer, a Varian electromagnet and an Oxford superconducting magnet. The NMR spectra were measured in the external magnetic fields $B_0$ = 1.17 and 8.0196 T, respectively. The nanorod spectra were obtained from Fourier transformation of the spin echo with 16-phase cycling sequence, while the wide spanned spectra of bulk samples were recorded using a frequency-shifted and summed Fourier transform processing technique [17]. Thallium NMR shifts are given relative to an aqueous 0.002 mol dm$^{-3}$ solution of $TlNO_3$, the position of which is assigned to the value of 0 ppm. In the external field $B_0$ = 8.0196 T, it shows the $^{205}Tl$ resonance at 196.9360 MHz. $^{205}Tl$ spin-lattice relaxation time $T_1$ was measured in the external magnetic field $B_0$ = 8.0196 T by means of a saturation comb pulse sequence.

$^{205}Tl$ NMR spectra of bulk and nanorod samples in $B_0$ = 8.0196 T are shown in Fig. 1. The spectra correspond to a nearly axially symmetric shielding tensor, in accordance with the tetragonal structure of the compound. At T = 291 K, the center of gravity of the bulk sample spectrum is located at $\sigma_{iso}$ = 3328 ppm (the precision is about 5 %), while the nanorod sample shows $\sigma_{iso}$ = 1124 ppm. One can find that the line positions and shapes of the spectra of bulk and

nanorod samples are substantively different. While Tl$_2$Se nanorods show a regular chemical (orbital) shift, bulk Tl$_2$Se exhibits a noticeable high frequency Knight shift being characteristic of conductors. The latter originates from the hyperfine interaction between nuclear spins and conduction electrons and is due primarily to *s* electronic orbitals having a nonzero value $|\Psi(0)|^2$ at the nucleus site. We note that the Knight shift in Tl metal is as large as 1.6% of the Larmor frequency [1,18], while the shift in bulk Tl$_2$Se is smaller: the sum of Knight and orbital shifts is 0.33%. It seems to be characteristic of a semimetal showing reduced density of states (DOS) at the Fermi level. This conclusion is supported by spin-lattice relaxation data presented below.

Low field ($B_0$ = 1.17 T) $^{205}$Tl and $^{203}$Tl spectra of nanorods (Fig. 2) are represented by single, nearly symmetric resonances with the second moment ($S_2$) values of 540 and 1090 kHz$^2$ for $^{205}$Tl and $^{203}$Tl isotopes, respectively. These values are much larger than those resulting from the contributions of the dipole-dipole interactions of nuclear spins and chemical shielding anisotropy. The former was estimated from the crystal structure of Tl$_2$Se using Van Vleck's formula [19] as $S_{2dd}$ ~ 2.5 kHz$^2$. The latter was evaluated [1] using the value of shielding anisotropy determined from the high field spectrum of nanorods and was found to be $S_2(\Delta\sigma)$ = 23 kHz$^2$ at $\nu_0$ = 28.6 MHz, corresponding to the Larmor frequency of $^{205}$Tl isotope in the magnetic field $B_0$ = 1.17 T. The large $S_2$ values observed imply that the line shape is related to essential indirect nuclear exchange interactions among the Tl nuclei that are characteristic for most of Tl-containing compounds [1-13]. Van Vleck has shown [19] that in the exchange-coupled nuclear systems, described by the Hamiltonian $\hat{H}=J_{jk}I_jI_k$, only exchange interaction between *unlike* spins contributes to the second moment of the NMR spectrum. Therefore in a crystal that contains two different isotopes, the ratio of their second moments is inversely proportional to the ratio of the isotope contents. For thallium, with natural abundances *f* = 29.5% for $^{203}$Tl and (1-*f*) = 70.5% for $^{205}$Tl, it should yield $S_2(Tl^{203})/S_2(Tl^{205})$ = (1-*f*)/*f* = 2.39, which is comparable with the experimental value. In the case in question, the second moment includes contributions of the exchange coupling, shielding anisotropy and dipolar contribution and may be written as

$$S_2 = \frac{1}{4}\sum_i J_{Tl-Tl}^2 + S_2(\Delta\sigma) + S_{2dd} \qquad (1)$$

(here we consider the scalar exchange term only). Since $S_2(\Delta\sigma)$ and $S_{2dd}$ contributions to $S_2$ are nearly the same for both thallium isotopes, the value of $J_{Tl-Tl}$ may be obtained from the difference of the second moments of $^{203}$Tl and $^{205}$Tl that suppresses the other contributions. Taking into account that each Tl atom has 12 nearest Tl neighbors, we found $J_{Tl-Tl}$ = 21 kHz that is characteristic of thallium compounds [1-13].

In bulk Tl$_2$Se, the signals of two isotopes overlap due to the large line widths, and their second moments are hardly determined. The larger line width in bulk Tl$_2$Se indicates larger exchange interaction in comparison with that in nanorods.

The spin-lattice relaxation time $T_1$ in bulk Tl$_2$Se is very short (of several milliseconds) that is characteristic of conductors. Fig. 3 shows $^{205}$Tl spin-lattice relaxation rate, $1/T_1$, measured as a function of temperature. One can find that at low temperatures $1/T_1 \sim T$, being a characteristic signature of the Korringa relaxation process that arises from the interaction of nuclear spins with conduction electrons. From linear fitting of the data below 200 K, we obtained $1/T_{1K}T$ = 1.2 s$^{-1}$K$^{-1}$, where the subscript $K$ denotes the Korringa process. The obtained value is much smaller than that of Tl metal, $1/T_1T$ =330 s$^{-1}$K$^{-1}$, which was roughly estimated [1,18] from the Korringa relation. Normally, the spin contribution to $1/T_{1K}T$ is proportional to the DOS at the Fermi-level. The obtained value of $1/T_1T$ indicates much smaller DOS at the Fermi level in dithallium selenide in comparison with that in metallic thallium.

Above 200 K, $1/T_1$ rises rapidly with an activation-like temperature dependence. Such behavior is characteristic of semiconductors [20], which show an increase in the relaxation rate due to increase in the number of carriers because of their thermal excitation across the energy gap. One can reconcile metallic behavior observed at low temperatures with the high-temperature semiconducting behavior by assuming a two-band model, with one band overlapping the Fermi level while the second band is separated from the Fermi level by an energy gap ΔE. In this case, the relaxation rate is given by the expression

$$\frac{1}{T_1} = \frac{1}{T_{1K}} + CT\exp(-\frac{\Delta E}{2k_B T}) \qquad (2)$$

where $1/TT_{1K}$ is the Korringa value obtained above, and the second term [20,21] is due to the band edge separated from the Fermi level. Here $C$ is a constant that depends on the effective mass and some other factors. We fixed the $1/TT_{1K}$ value obtained from the low-temperature fit and found that Eq. 2 yields good agreement with the experimental data. The energy gap found from the least-squares fit is ΔE = 0.16 eV.

In $Tl_2Se$ nanorods, $T_1$ varies from 3.9 s at 70 K to 40 ms at 291 K (Fig. 4) and shows an activation behavior characteristic of semiconductors. The activation energy was determined to be ΔE =0.24 eV.

Insufficiently large (comparing with Tl metal) values of the Knight shift and Korringa term $1/TT_{1K}$ in bulk $Tl_2Se$, along with the unusual temperature dependence of the spin-lattice relaxation, allows one to conclude that this compound is a semimetal and shows significant reduction in the DOS near the Fermi level (so called pseudogap). The Fermi level is likely located in the pseudogap, thus the main contribution to the high temperature conductivity comes from the delocalized carriers due to their excitation across the pseudogap, though in the absence of the real forbidden band.

The semiconductor behavior of $Tl_2Se$ nanorods may be explained in terms of the semimetal-semiconductor transformation due to two-dimensional quantum confinement on the resulting quasi-one-dimensional electron transport properties. Quantum confinement is observed when the sample size becomes smaller than the de Broglie wavelength. In metals, the de Broglie wavelength of the carriers is of the order of the lattice spacing. Thus quantum confinement effects may be visible only for very small particles. However, the situation is different for semimetals and semiconductors, in which the de Broglie wavelength may exceed the lattice spacing by several orders of magnitude. For example, the semimetal bismuth, which is characterized by a low carrier concentration, small effective mass, long mean free path of carriers and a highly anisotropic Fermi surface, exhibits quantum confinement effects at a thickness of ~1000 Å [22]. Lutskii [23] and Sandomirskii [24] have shown that dropping the top of the valence band and lifting the bottom of the conduction band with reduced particle size may result in appearance of the energy gap, i.e. in metal-isolator transition. Recent experiments [25-27] provide an indication of the semimetal-semiconductor transition in Bi nanowires as the wire diameter is reduced. A similar transformation is likely observed in the compound under study. Our experimental data on the Knight shift and Korringa term are in accordance with a low DOS near the Fermi level and a low carrier concentration that are required for the semimetal behavior. At that, $Tl_2Se$ nanorods are suggested to exhibit rather long de Broglie wavelength and mean free path of carriers that should exceed the average diameter of the nanorods, 75 nm, in order to guarantee the quantum confinement of carriers inside the cylindrical potential well.

One of the authors (A.M.P.) thanks Y. Chaikovsky for helpful discussions. A part of this work is supported by the National Natural Science Foundation of China (#20571001).

**Figure captions**

Fig. 1. $^{205}$Tl NMR spectra of bulk and nanorod samples of Tl$_2$Se at T=291 K and B$_0$=8.0196 T.

Fig. 2. $^{205}$Tl and $^{203}$Tl NMR spectra of bulk and nanorod samples of Tl$_2$Se at T=291 K and B$_0$=1.17 T.

Fig. 3. Temperature dependence of $^{205}$Tl spin-lattice relaxation rate of Tl$_2$Se bulk sample in B$_0$=8.0196 T. Linear fit at low temperatures is shown by dotted line.

Fig. 4. Temperature dependence of $^{205}$Tl spin-lattice relaxation rate of Tl$_2$Se nanorod sample in B$_0$=8.0196 T. Exponential fit is shown by dotted line.

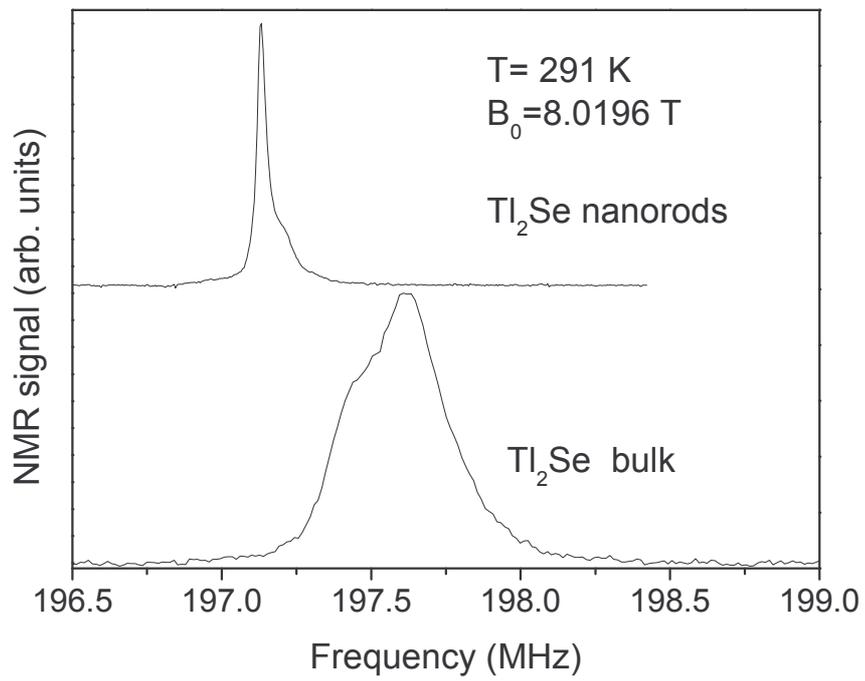

Fig. 1.

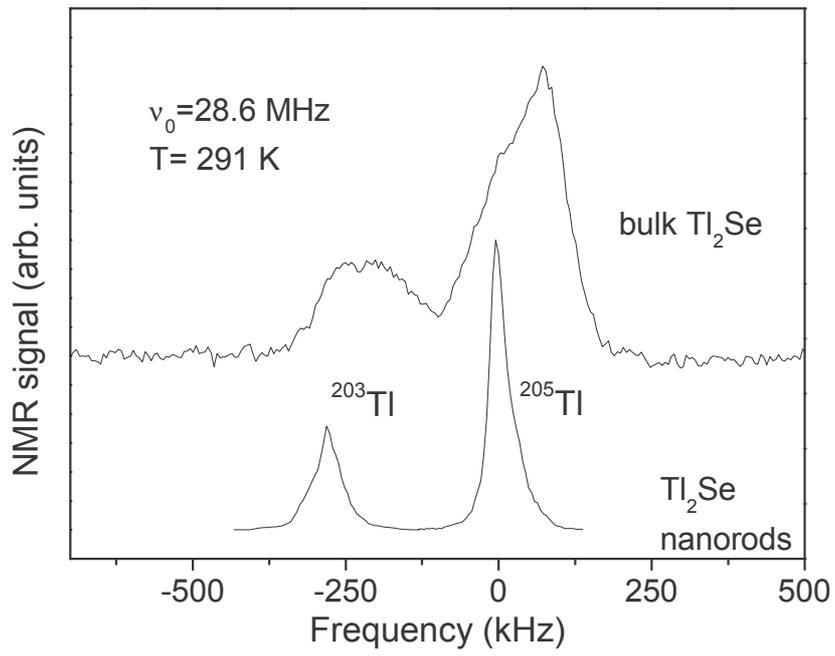

Fig. 2.

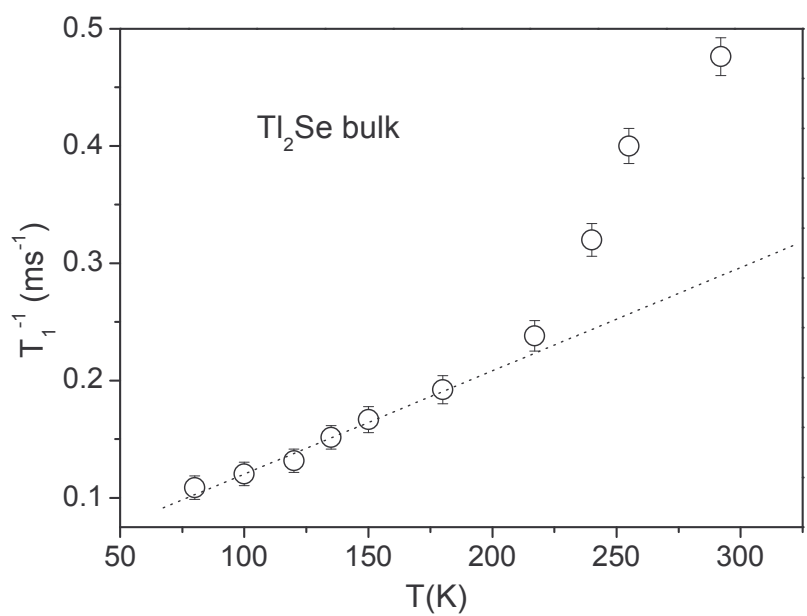

Fig. 3.

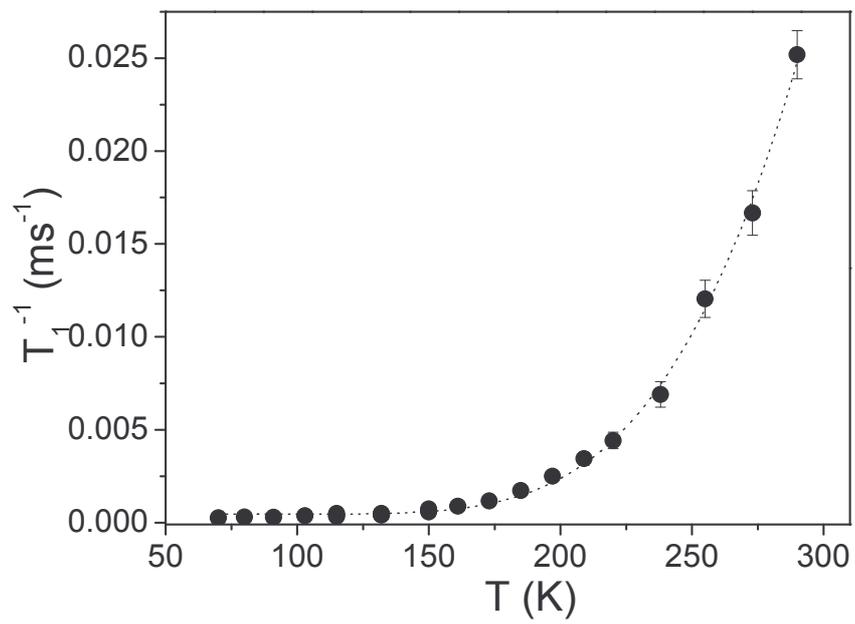

Fig. 4.